\documentclass[reprint,aps,ams,amsmath,prx,twocolumn,longbibliography,superscriptaddress]{revtex4-1}
\usepackage{graphics}
\usepackage{graphicx}
\usepackage{epstopdf, epsfig}
\usepackage{amsmath}
\usepackage{amssymb}
\usepackage{amsfonts}
\usepackage{bm}
\usepackage{color}
\usepackage{bbm}
\usepackage{hyperref}
\usepackage[normalem]{ulem}
\usepackage{comment}
\usepackage{soul}

\newcommand{\be}{\begin{equation}}
\newcommand{\ee}{\end{equation}}
\def\rr#1{(\ref{#1})}

\begin{document}

\title{Wave-crests around obstacles in odd viscous liquids}

\author{E. Kirkinis}
\affiliation{Center for Computation and Theory of Soft Materials, Robert R. McCormick School of Engineering and Applied Science, Northwestern University, Evanston IL 60208 USA}

\author{A. Levchenko}
\affiliation{Department of Physics, University of Wisconsin-Madison, Madison, Wisconsin 53706, USA}

\date{\today}

\begin{abstract} 
The values of liquid odd-viscosity coefficients remain largely unknown, with only a single experimental measurement reported to date [Nature Physics \textbf{15} (11), 1188–1194 (2019)]. In this work, inspired by the well-known consequences of dispersion surfaces in classical liquids from the work of Lighthill, we theoretically determine the shapes of constant-phase wave crests formed around obstacles moving at constant velocity in two- and three-dimensional odd viscous liquids, which may or may not undergo rigid rotation. From this analysis, we derive parametric relations that the odd-viscosity coefficients must satisfy, providing a framework for their experimental determination.
\end{abstract}

\maketitle


In two dimensional incompressible liquids, effects of odd viscosity
can only be met at a free surface \cite{Ganeshan2017, *Kirkinis2022a}. In such a configuration the Irvine group at Chicago observed experimentally the attenuation of surface 
undulations in blobs of spinning magnets caused by the presence of odd viscosity, and 
measured its value \cite{Soni2019}. In three-dimensional incompressible or two dimensional compressible odd viscous liquids however, the values of the odd viscosity coefficients still remain undetermined
\footnote{Odd viscosity coefficients for gas flow under a magnetic field were measured in a series of experiments 
by Beenakker and coworkers \citep{Hulsman1970}}.

\begin{figure*}[t!]
\includegraphics[height=2.2in,width=7.5in]{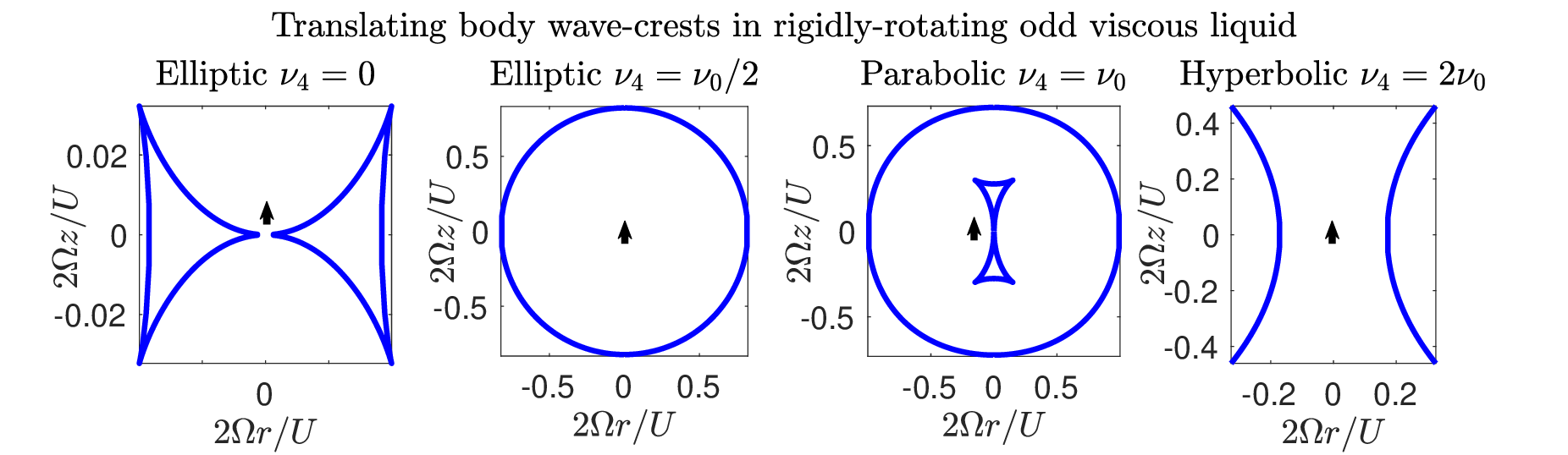}
\vspace{-10pt}
\caption{Wave-crests of constant phase (Eq. \rr{x} with dispersion equation \rr{P}, see Appendix for the parametric representation of the curves) around a body moving with velocity $U \hat{\mathbf{z}}$, in the frame rotating with angular velocity $\Omega \hat{\mathbf{z}}$ of a three-dimensional odd viscous liquid characterized by the constitutive law \rr{sigma1}. Arrows denote the direction of motion of the body and $r = (x^2 + y^2)^{1/2}$. Compare with the wave-crests in figure \ref{dispersion_surfaces_non_odd} determined from the same theory for a classical (non-odd) Newtonian liquid, which agreed exceedingly well with observation and experiment. Curve parametric forms are delegated to Appendix
\ref{sec: parametric}.  
\label{wave_crests_odd}}
\end{figure*}

In this Letter, we establish a theoretical protocol for determining the values of odd viscosity coefficients in the two configurations discussed above. This can be realized experimentally by observing the shapes of constant-phase wave crests generated by a translating or oscillating body in a possibly rigidly rotating odd viscous liquid. The wave-crest shapes (cf. Fig. \ref{wave_crests_odd}) can be obtained directly from the dispersion relation of the unforced system by accounting for the Doppler shift in the forcing frequency (cf. \citealt{Lighthill1967} and Appendix \ref{sec: dispersion}). In classical fluid mechanics, this theory has been shown to agree remarkably well with experiment -- for example, it accurately predicts the herringbone pattern of internal gravity waves formed behind an ascending sphere in a density-stratified fluid \cite{Mowbray1967}, as well as the Kelvin wake generated by a moving ship \cite{Lighthill1978} (see our summary in Fig. \ref{dispersion_surfaces_non_odd} in Appendix \ref{sec: dispersion}). This formulation is valid at distances much larger than the size of the obstacle.

We consider a three-dimensional \emph{incompressible} odd viscous liquid, subject to the constitutive law \rr{sigma1}, rigidly rotating with angular velocity $\Omega$ in the $z$-direction. The vorticity equation (the curl of the time-dependent Stokes equations, cf. \cite[Eq. (14.4)]{Landau1987}) 
\be \label{vort1}
\partial_t\textrm{curl} \mathbf{v} = -(\widehat{\mathcal{S}} - 2\Omega) \frac{\partial \mathbf{v}}{\partial z}
\ee
where 
\be \label{Leta4}
\widehat{\mathcal{S}} = (\nu_o-\nu_4)\nabla^2_2 + \nu_4\partial_z^2, 
\ee
is a differential operator that can be elliptic, parabolic or hyperbolic, 
$(\nu_o,\nu_4) = (\eta_o,\eta_4)/\rho$, $\rho$ is the constant density of the liquid and $\nabla^2_2 = \partial_x^2 + \partial_y^2$, 
gives rise to the dispersion equation
$P(\omega, \mathbf{k}) \equiv \omega^2 k^2 - 
\left(2\Omega -  \mathcal{S}(\mathbf{k})\right)^2k_z^2 = 0 $,
where
$
\mathcal{S}(\mathbf{k}) = - (\nu_o - \nu_4) (k_x^2+k_y^2) - \nu_4 k_z^2. 
$
When a forcing effect, generated by the presence of an obstacle, of the form 
\be \label{tf}
e^{-i\omega_0 t} f(\mathbf{x} - \mathbf{U} t),
\ee
is superposed on a limited region of the liquid described by Eq. \rr{vort1}, translating with velocity $\mathbf{U}$ and oscillating with frequency $\omega_0$,
the wavenumber $\mathbf{k}$ satisfies a dispersion equation
\be \label{P}
P(\omega_0 +\mathbf{U}\cdot \mathbf{k}, \mathbf{k}) \equiv (\omega_0 + \mathbf{U}\cdot \mathbf{k})^2 k^2 - 
\left(2\Omega -  \mathcal{S}(\mathbf{k})\right)^2k_z^2 = 0 
\ee
cf. Appendix \ref{sec: dispersion} and \citet{Lighthill1967}, representing a surface in wavenumber space (a gallery of such surfaces for Eq.  \rr{P} is delegated to Supplementary Information figure S-I). 

The shapes of wave-crests of constant phase are then determined by the reciprocal polar relation to \rr{P}, that is the trace of the curve
\be \label{x}
\mathbf{x} = A \frac{\nabla_\mathbf{k} P(\omega_0 +\mathbf{U}\cdot \mathbf{k}, \mathbf{k}) }{(\mathbf{k} \cdot \nabla_\mathbf{k}) P(\omega_0 +\mathbf{U}\cdot \mathbf{k}, \mathbf{k}) },
\ee
where $A$ is a constant, see Appendix \ref{sec: dispersion} for a derivation of Eq. \rr{x}. It is clear that the curves \rr{x} denote wave fronts of constant phase since $\mathbf{k} \cdot \mathbf{x} = A$.  
In figure \ref{wave_crests_odd} we display the curves \rr{x} (surfaces of revolution), formed around a body moving with velocity $\mathbf{U} = U\hat{\mathbf{z}}$
(in the absence of oscillation ($\omega_0\equiv 0$)),
in the three-dimensional odd viscous liquid described by Eq. \rr{vort1}. The arrow denotes the direction of the obstacle motion.
The curves \rr{x} depend on a single dimensionless number
\be \label{R}
R = \frac{2\Omega \nu_0}{U^2},
\ee
which can be interpreted as the ratio of the Taylor number $\frac{2\Omega L}{U}$ to the odd Reynolds number 
$ \frac{ UL}{\nu_0}$, where $L$ is a characteristic length. Each of the four panels in figure \ref{wave_crests_odd}, corresponds to a specific case of interest to  experiment~ \footnote{Depending on the relative values of the viscosity coefficients, the differential operator \rr{Leta4} falls into three distinct classes (assuming $\nu_4 > 0$ for simplicity): it is elliptic when $\nu_o>\nu_4$,
hyperbolic when $\nu_o<\nu_4$, and parabolic when $\nu_o=\nu_4$.}. The leftmost, elliptic panel with $\nu_4 \equiv 0$ represents a liquid whose constitutive law resembles that of a two-dimensional odd viscous fluid. The resulting surfaces exhibit cusps similar to those trailing a ship (cf. left panel of Fig.~\ref{dispersion_surfaces_non_odd}); however, in this case, the cusps appear laterally.

The second panel from the left $\nu_4 = \nu_0/2$ is a case of interest in the current literature \cite{Markovich2021} and it resembles values taken in the experimental system of Beenakker and coworkers, cf. \cite{Hulsman1970} for a gas in a magnetic field. Expressing $\widehat{\mathcal{S}}$ in the form $(\nu_o-2\nu_4)\nabla_2^2 + \nu_4 \nabla^2$ shows that in this case it reduces to the
Laplace operator $\nu_4 \nabla^2$. The surfaces resemble those of non-odd viscous liquids (right-most panel of Fig. \ref{dispersion_surfaces_non_odd}).

The parabolic case, $\nu_4 = \nu_o$, is also of interest, as the operator in \rr{Leta4} varies only along the axis of rotation. As in the previous case, there exists a spherical surface which now, however, envelops two cusp-like surfaces -- one ahead of and one trailing the body.

Finally, the rightmost hyperbolic panel is of interest as it has been shown that characteristics in this case are oblique \cite{Kirkinis2024} and lie on a local ray or Monge cone \cite[p. 601]{Courant1962}. Here the wave-crests form a hyperboloid of revolution around the body. The parametric form of \rr{x} for each panel of figure \ref{wave_crests_odd} is delegated to Appendix
\ref{sec: parametric}. 

\begin{figure}[t!]
\includegraphics[height=2.6in,width=3.4in]{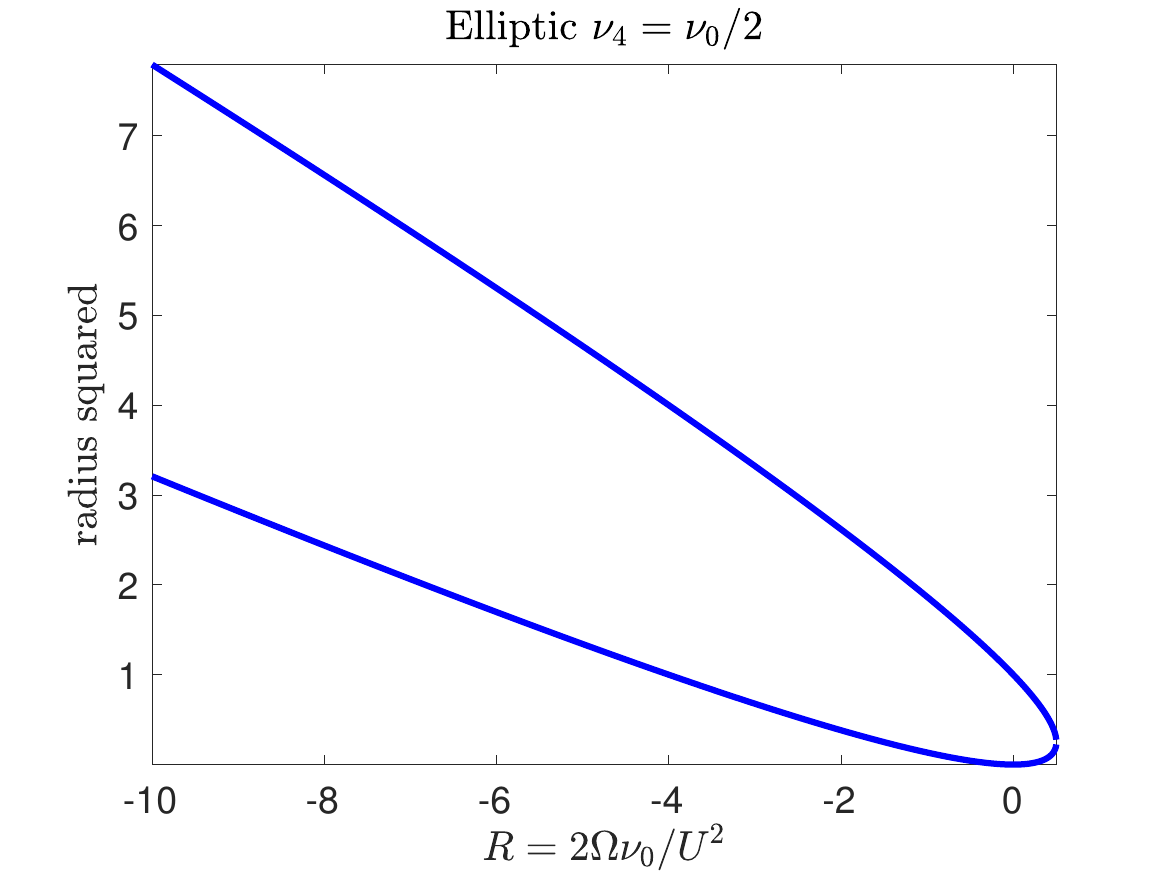}
\vspace{-10pt}
\caption{Squared radius \rr{radius2} of spherical fronts (second panel from the left in figure \ref{wave_crests_odd})
for the determination of the value of the odd viscosity coefficients in an experiment. 
\label{radius2R}}
\end{figure}

To determine the value of the odd viscosity coefficient in an experiment, consider the elliptic case $\nu_4  = \nu_0/2$ of the second panel from the left in figure \ref{wave_crests_odd}. The wave-fronts are circles (spheres) whose radius squared 
\be \label{radius2}
\frac{2 \left(2 R -1\right) R^{2} \left(R \mp \sqrt{-2 R +1}-1\right)}{\left(\sqrt{-2 R +1}\mp2 R +1\right)^{2} \left(\sqrt{-2 R +1}\pm1\right)^{2}}
\ee
is a simple function of the dimensionless number $R$ in \rr{R}. 
We display the squared radius \rr{radius2} in figure \ref{radius2R} whose lower branch (upper sign in \rr{radius2}) corresponds to the spherical front displayed in figure  \ref{wave_crests_odd}. It is clear that by observing the evolution of the radius of the front in an experiment, values for the 
number $R$ in \rr{R} can be determined from  figure \ref{radius2R} and thus the value of the odd viscosity coefficient $\nu_0$ can likewise be determined. 

 \begin{figure}[t!]
\includegraphics[height=2.6in,width=3.4in]{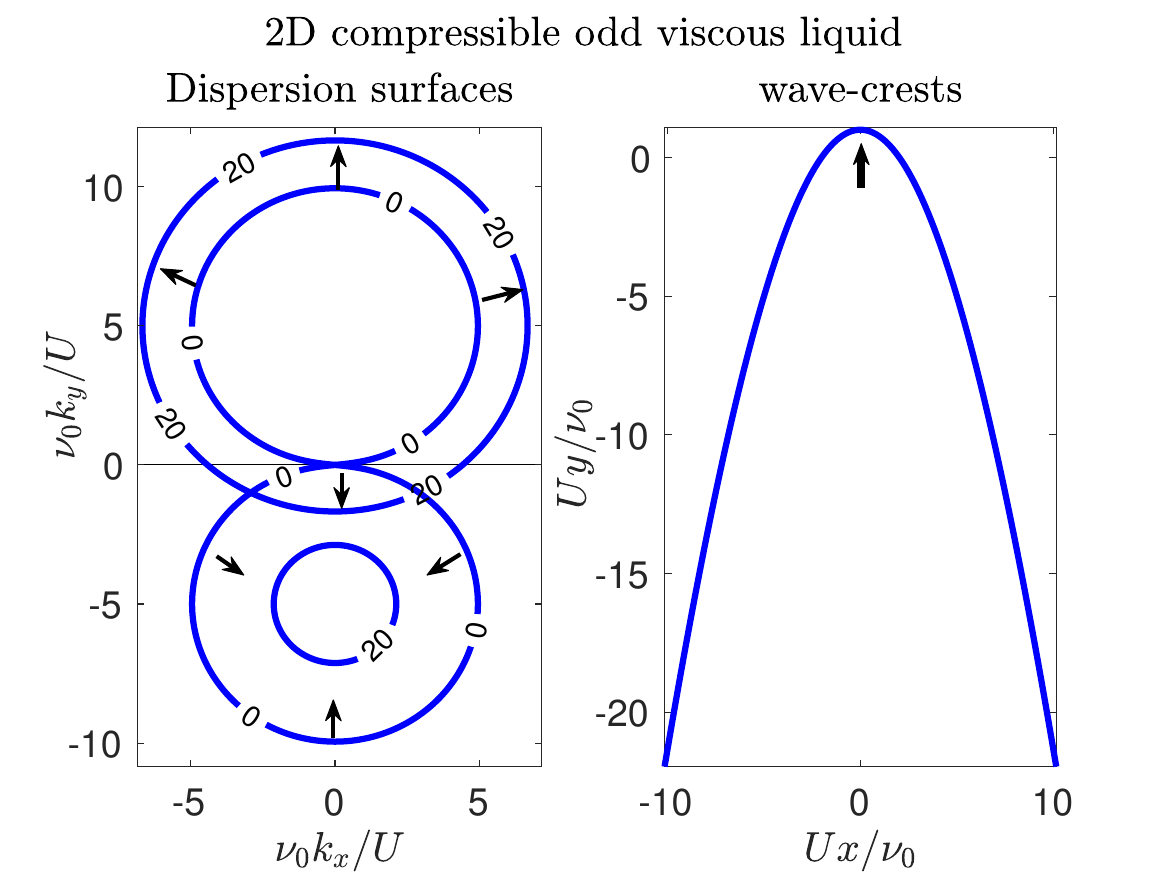}
\vspace{-10pt}
\caption{Left panel: Dispersion surfaces \rr{P2D} where the number on the curve denotes the forcing frequency $\omega_0$ and the thin arrows denote the direction of increasing $\omega_0$. Thus long waves trail behind the obstacle and away from its path while short waves point in all directions and are concentrated close to the obstacle's path. Right panel: 
Wave-crests of constant phase (Eq. \rr{x} with dispersion equation \rr{P2D}, see Appendix for the parametric representation of the curves). Thick arrow denotes the direction of motion of the traveling force effect (whose frequency $\omega_0$ has been set to zero).
\label{wave_crests_odd2D}}
\end{figure}

In the absence of rotation ($\Omega \equiv 0$), the parametric expressions for \rr{x} simplify significantly. For instance, in the elliptic $\nu_4=\nu_0/2$ case 
\rr{x} becomes
$r = \sqrt{-s^{2}+4}/4$ and $z = s/4$
where $s = \nu_0 k_z /U$ and the coordinates have been nondimensionalized by $(r, z) \rightarrow \nu_0 (r,z) / U$. This is a circle (sphere) of radius $1/2$ (thus similar to the second panel from the left in figure \ref{wave_crests_odd}), and the value of $\nu_0$ can be determined in an experiment by comparing the front sizes as the velocity $U$ of the moving body varies. 

In the parabolic $\nu_4=\nu_0$ case \rr{x} becomes
$r = \sqrt{(s^{2}-1)/s^{6}}$ and $z = (2 s^{2}-1)/s^{3}$. These fronts are similar to those at the center of the second panel from the right in figure \ref{wave_crests_odd}), and the value of $\nu_0$ can likewise be determined as the experimentalist adjusts the velocity $U$ of the obstacle and compares the resulting front sizes.

We proceed with an investigation of wave crests around an obstacle (forcing effect \rr{tf}) moving with constant velocity $U$ in the $y$-direction, in a two dimensional \emph{compressible} odd viscous liquid. 
The constitutive law of such a liquid consists of the upper left two-by-two block of the first matrix in \rr{sigma1}, obtained from expression \rr{sigma2}. The equations of motion have been variously derived in the literature
with the most succinct form being $\partial_t^2 \chi + \left[ (kc)^2 + (\nu_0k^2)^2 \right]\chi =0$, where $\chi = \mathbf{k} \cdot \mathbf{v}$ and $c$ is the speed of sound \cite[\S89]{Lifshitz1980}. Thus, the dispersion equation reads
$P(\omega, \mathbf{k}) \equiv \omega^2 -  (kc)^2 - (\nu_0k^2)^2=0$, so that the traveling forcing effect \rr{tf} gives rise to 
\be \label{P2D}
P(\omega_0 +\mathbf{U}\cdot \mathbf{k}, \mathbf{k}) \equiv (\omega_0 + Uk_y)^2 -  (kc)^2 - (\nu_0k^2)^2=0
\ee
that is to be employed in the determination of the wave-crests \rr{x}. 

On the left panel of figure \ref{wave_crests_odd2D} we display the dispersion surfaces \rr{P2D} where the number on the curve denotes the forcing frequency $\omega_0$. The thin black arrows denote the direction of increasing $\omega_0$. Thus, long waves (small $k$) trail the rising obstacle and are found away from its path. On the other hand, short waves (large $k$) are found in all directions and are close to the path of the obstacle. This behavior is displayed on the left panel of figure \ref{wave_crests_odd2D} where we plot the 
wave-crests \rr{x}, whose parametric form is delegated to the Appendix \ref{sec: dispersion}, and the thick black arrow denotes the direction of motion of the translating forcing effect in the absence of an oscillation ($\omega_0\equiv 0$). The curves depend on a single dimensionless number
\be \label{V}
V = \frac{U}{c},
\ee
that is, the ratio of the obstacle velocity to the speed of sound. 

Patterns similar to those shown in Fig.~\ref{wave_crests_odd} can be derived for a variety of related systems -- for instance, for wave crests generated by a body that both oscillates and translates. In the context of classical liquids, formulations based on dispersion surfaces have again shown excellent agreement between theory and experiment \cite{Mowbray1967a, Stevenson1969}. Other experimental tests to determine the odd viscosity coefficients can also be envisioned -- for example, by considering the steady motion of a body \cite{Peat1976} or its acceleration \cite{Woodhead1983} in a simultaneously rotating and stratified liquid.
Ref. \cite{Kirkinis2024} provides an alternative test of determining the odd viscosity coefficients, that of observing the precession rate of prograde or retrograde waves with respect to the rotating frame of reference. 

The presence of shear viscosity (not investigated in this Letter by assuming it is subdominant to odd viscosity) may lead to some dissipation of the patterns established here.
As shown however for systems endowed with both dispersion \emph{and} dissipation,  
\cite[p. 254]{Zauderer1983} and
\cite{Sirovich1970,*Sirovich1971}, there is a diffusion effect appearing in the wake of characteristics which will yield the major contribution at long times. Thus, although the fronts will die-out as $t\rightarrow \infty$, the decay of the diffusion effect will only be algebraic rather than exponential.  

In this Letter, we have established the shapes of wave fronts corresponding to all wavenumbers. When, however, $k_z$ is close to zero, the theory predicts special effects such as the Taylor-Proudman theorem, Taylor columns, and Long waves \cite{Lighthill1967}. In the context of odd viscous liquids, these correspond to structures that form ahead of and behind a slowly moving body, under the assumption that $\nu_o \gg U L$, where $L$ is a characteristic length (e.g., the size of the body; cf. \cite{Kirkinis2023taylor,*kirkinis2023halos}). An extensive discussion of these effects from the viewpoint of dispersion surfaces is provided in the Supplementary Information addendum.

\section*{Acknowledgements}

This work was supported by National Science Foundation Grant No. DMR-2452658 and H. I. Romnes Faculty Fellowship provided by the University of Wisconsin-Madison Office of the Vice Chancellor for Research and Graduate Education with funding from the Wisconsin Alumni Research Foundation.

\section*{Data availability} 

All data presented in the figures were generated from analytical expressions derived and defined in the paper. The Wolfram Mathematica code used to produce the plots will be made available by the authors upon reasonable request.

\appendix

\section{\label{sec: constitutive}Odd constitutive law}

When a liquid is endowed with a preferred direction, the Newtonian constitutive law admits many more terms. These were 
succinctly expressed by Lifshitz \& Pitaevskii \cite[\S 13]{Landau1981}. Among them, 
there are terms which are nondissipative, in the sense that they do not contribute to
viscous heating in the energy conservation law. These extra terms have the form
\begin{align}
\sigma_{\alpha\beta}'& = \eta_3\left[ V_{\beta\gamma}  b_{\alpha\gamma}
+V_{\alpha\gamma}   b_{\beta\gamma} \right] \nonumber\\
&+ V_{\gamma\delta} 
 (2\eta_4-\eta_3)(b_{\alpha\gamma} b_\beta b_\delta + b_{\beta\gamma} b_\alpha b_\delta) , \label{seta1234} 
\end{align}
where $b_{\alpha\beta} = \epsilon_{\alpha\beta\gamma}b_\gamma$ and $b_\gamma$ is the vector denoting the preferred direction, as to be discussed in more detail below. $\eta_i$ are viscosity coefficients and $\epsilon_{\alpha\beta\gamma}$ is the alternating
tensor, summation is implied on repeated indices and $V_{\gamma\delta} = \frac{1}{2}\left( \frac{\partial u_\gamma}{\partial x_\delta} + \frac{\partial u_\delta}{\partial x_\gamma}\right)$ is the rate-of-strain tensor. 
Thus, a general form for the 
Cauchy stress of a classical liquid that contains the terms \rr{seta1234} is  
\be \label{sabcd}
\sigma_{\alpha\beta} = \eta_{\alpha \beta \gamma \delta} V_{\gamma \delta}
\ee
and, as can easily be verified, the part of the tensor $\eta_{\alpha \beta \gamma \delta}$ associated
with the terms \rr{seta1234} is antisymmetric with respect to the exchange of the first and second 
pairs of indices:
$
\eta_{\alpha \beta \gamma \delta} = -\eta_{\gamma \delta\alpha \beta }. 
$
Thus a liquid whose constitutive law depends on external fields, say $\mathbf{b}$, that 
change sign under time-reversal, satisfies the symmetry of the kinetic
coefficients $\eta_{\alpha\beta\gamma\delta}$ (or Onsager principle, \cite[\S 13]{Landau1981}) when
\be
\eta_{\alpha\beta\gamma\delta}(\mathbf{b})
= \eta_{\gamma\delta\alpha\beta}(-\mathbf{b}).
\ee
This type of behavior can be induced, for instance, by a magnetic field, giving rise to an
\emph{anisotropy axis} along its direction.

We consider an odd viscous liquid whose direction of anisotropy is the $z$-axis.
In this case the constitutive law \rr{seta1234} can be expressed in more familiar terms
in Cartesian coordinates
\begin{widetext}
\be \label{sigma1}
\bm{\sigma}' = \eta_o 
\left(\begin{array}{ccc}
-\left(\partial_x v + \partial_y u \right) &  \partial_x u  - \partial_y v & 0\\
 \partial_x u  - \partial_y v & \partial_x v + \partial_y u   & 0\\
0&0&0
\end{array}
\right) +
\eta_4 
\left(\begin{array}{ccc}
0 & 0  & -(\partial_y w + \partial_z v)\\
0 & 0  & \partial_x w + \partial_z u\\
 -(\partial_y w + \partial_z v)&\partial_x w + \partial_z u&0
\end{array}
\right).
\ee
\end{widetext}
Either of these tensors signifies the stress response in a \emph{three dimensional} odd viscous liquid
described by a velocity field $\mathbf{v} = u \hat{\mathbf{x}} + v \hat{\mathbf{y}} +w \hat{\mathbf{z}} $
which satisfies the isochoric constraint
\be \label{inc}
 \partial_x u + \partial_y v + \partial_z w = 0.
\ee 
Here, we employed the notation $\eta_o$ instead of $\eta_3$ and we note that the sign in front of the odd viscosity coefficients is determined by the direction of the field $\mathbf{b}$.

For a two dimensional liquid, the expression for the odd viscosity dependence is adopted from the literature
\cite{Avron1998,Lapa2014} 
\begin{align}  \label{sigma2}
\sigma^o_{ik}& = - \eta^o (\delta_{i1}\delta_{k1} - \delta_{i3}\delta_{k3} ) 
\left( \frac{\partial u_1}{\partial x_3} +\frac{\partial u_3}{\partial x_1}  \right) \nonumber\\
&+ \eta^o \left( \delta_{i1}\delta_{k3} + \delta_{i3} \delta_{k1}    \right) 
\left( \frac{\partial u_1}{\partial x_1} - \frac{\partial u_3}{\partial x_3}\right).
\end{align}

 \begin{figure}[t!]
\includegraphics[height=1.6in,width=3.5in]{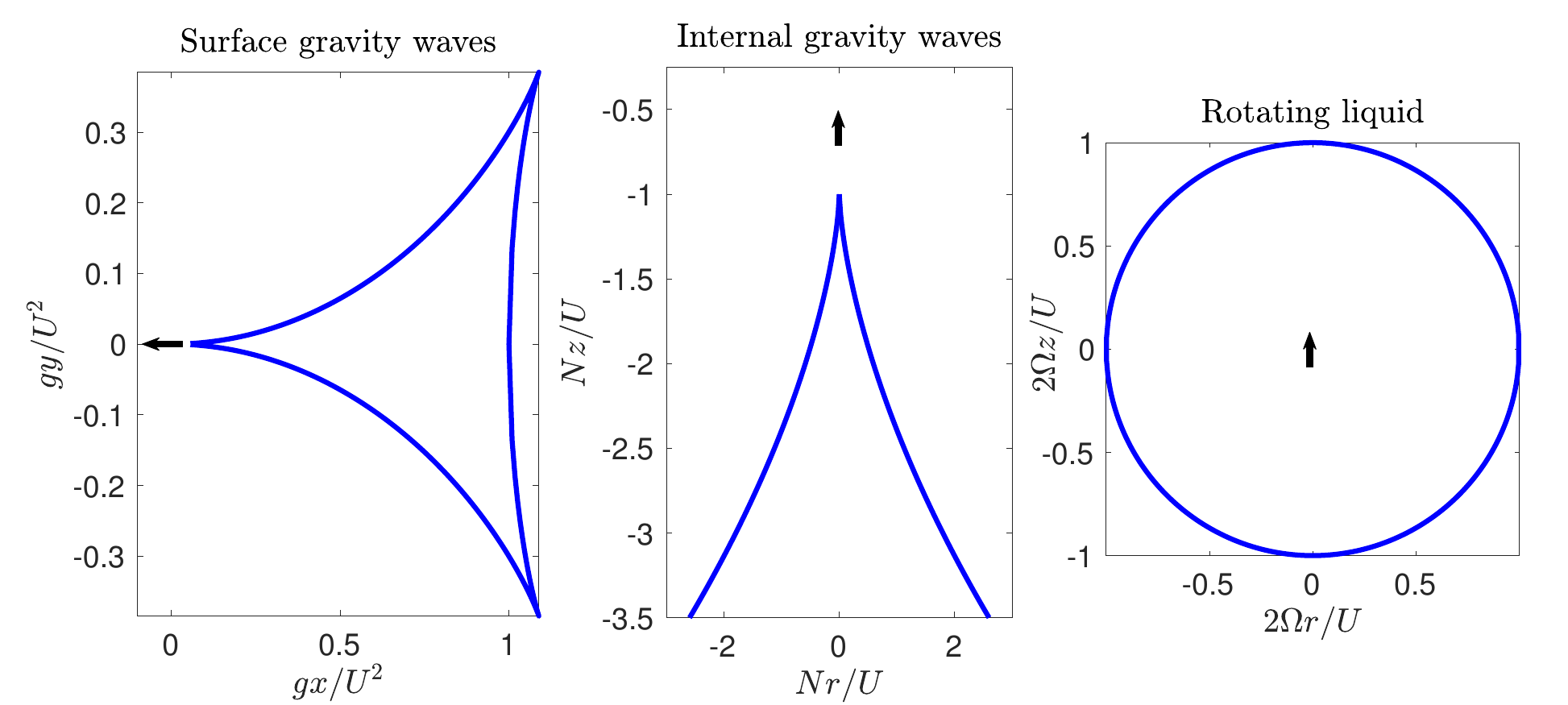}
\vspace{-10pt}
\caption{Known shapes of wave-crests of constant phase, formed behind an obstacle moving with constant velocity (black arrow) in a classical Newtonian liquid theoretically predicted by \citet{Lighthill1967,Lighthill1978}. Surface gravity waves observed to form behind ships (ship Kelvin waves \cite{Lighthill1978}). Internal gravity waves experimentally observed to form behind an ascending sphere in stratified liquid \cite{Mowbray1967}. The spherical wave-crests form around an obstacle moving with constant speed in a rigidly-rotating liquid. 
\label{dispersion_surfaces_non_odd}}
\end{figure}

\section{\label{sec: dispersion}General theory of dispersion surfaces for oscillatory forcing}
In this Appendix we describe the theory of dispersive waves far from a forcing region leading to the wave-crests parametric equation \rr{x}, by following \citet[section 2]{Lighthill1967}. 

We consider an oscillating forcing region moving with velocity $\mathbf{U}$ and having the form $e^{-i\omega_0 t} f(\mathbf{x} - \mathbf{U} t)$, being the inhomogeneous part of a partial differential equation (PDE)
and representing the action of external generalized forces. Such a PDE can have the form
\be \label{L7}
P\left( i\partial_t, -i \nabla\right) \phi = e^{-i\omega_0 t} f(\mathbf{x} - \mathbf{U} t)
\ee
where $P$ is a polynomial and $\phi = \phi_0 e^{i(\mathbf{k} \cdot \mathbf{x} - \omega t)}$. 
In the absence of the forcing term the dispersion relation reads 
\be \label{Papp}
P(\omega, \mathbf{k}) = 0. 
\ee

With $f$ being nonzero only inside a limited forcing region, it has a Fourier integral representation 
$f(\mathbf{x}) = \int d^3k  F (\mathbf{k}) e^{i\mathbf{k} \cdot \mathbf{x} }$. The solution of \rr{L7} is 
\be \label{L8}
\phi = \int d^3 k \frac{ F (\mathbf{k}) e^{i(\mathbf{k} \cdot ( \mathbf{x} - \mathbf{U} t) - \omega_0 t) }}{P(\omega_0 + \mathbf{U} \cdot \mathbf{k}, \mathbf{k})}.
\ee
The unique solution is determined by considering the Sommerfeld radiation condition whereby waves exist originating only from the forcing region and not from free waves. 
At distances large from the forcing region, the integral can be evaluated by employing stationary-phase methods developed by Lighthill \cite{Lighthill1960, *Lighthill1966}. At each point of $\mathbf{k}$ space we draw an arrow normal to the surface
\be \label{L9}
P(\omega_0 + \mathbf{U} \cdot \mathbf{k}, \mathbf{k})=0
\ee
 in the direction of increasing frequency $\omega_0 + \mathbf{U} \cdot \mathbf{k}$. This denotes the direction of waves stretching-out from the forcing region and it is illustrated in the left panel of figure \ref{wave_crests_odd2D}.
 
This theory has been very successful in predicting the shape of wave-crests behind or around obstacles in a classical (non-odd) liquid. We display three such characteristic cases in figure \ref{dispersion_surfaces_non_odd}. 

The simplest way to derive Eq. \rr{x} for the wave-crests is to consider \rr{Papp} as defining the collection of integral surfaces $\varphi = \varphi(\mathbf{x}, t)$. The normal vectors at the points $(\mathbf{x}, t, \varphi)$ have to satisfy \rr{Papp} which, in this sense, it generates a one-parameter family of tangent planes, determined by $\mathbf{k}$, to the integral surfaces at each point in the $(\mathbf{x}, t, \varphi)$ space. The intersection of the Monge cone (the cone enveloped by the tangent planes) with the integral surface determines the field of characteristic directions. 

Let $\mathbf{x}= \mathbf{x}(s)$, $t = t(s)$ and $\varphi=\varphi(s)$ be those intersections
The equations that define these curves (characteristic equations) are 
\be \label{char}
\frac{d\mathbf{x}}{ds} = \nabla_\mathbf{k} P, \frac{d\varphi}{ds} =( \mathbf{k} \cdot \nabla_\mathbf{k}) P,
\frac{d\mathbf{k}}{ds} = - \nabla_\mathbf{x} P-  \mathbf{k}\partial_\varphi P, 
\ee
$dt/ds = \partial_\omega P$ and $d\omega/ds = -\partial_t P - \omega \partial_\varphi P$
 (cf. \cite[\S 2.4 \& p. 688]{Zauderer1983} for a simple introduction and \cite{Courant1962} for a general theory). 
 
 For a $P$ having the form \rr{Papp}, characteristic equations \rr{char} simplify significantly. Eliminating $s$ from the first two equations in \rr{char} and integrating gives
 \be \label{x1}
\mathbf{x} =  \frac{\nabla_\mathbf{k} P(\omega, \mathbf{k}) }{(\mathbf{k} \cdot \nabla_\mathbf{k}) P(\omega, \mathbf{k}) } (\varphi - \varphi_0) + \mathbf{x}_0
\ee
where $\varphi(\mathbf{x_0}) = \varphi_0$. Thus, from \rr{x1} one obtains \rr{x} by considering the traveling forcing effect \rr{tf}, where a choice has been made to parameterize the characteristic rays $\mathbf{x}= \mathbf{x}(s)$ with respect to the wavenumber components rather than on time.

\section{\label{sec: parametric} Wave-crests parametric forms}
The wave crests \rr{x} are, in general, surfaces depending on two parameters. The preferred direction of the constitutive law \rr{sigma1} allows us to reduce them to single-parameter surfaces of revolution. This is possible because the dispersion equation \rr{P} can be solved for $\kappa = (k_x^2 + k_y^2)^{1/2}$ as a function of $k_z$. When substituted into \rr{x} this gives a single parameter curve (surface of revolution). The dimensionless parameter we employ below is 
\be
s = \frac{U k_z}{2\Omega},
\ee
and its domain is determined by the respective problem. The four cases displayed in figure \ref{wave_crests_odd} have the following parametric representation $\mathbf{x}(s) = (r(s), z(s))$ where the coordinate functions are 
\be
(r,z) \rightarrow  \frac{2\Omega }{U}(r,z).
\ee
Elliptic case $\nu_4 =0 $ (left-most panel in figure\ref{wave_crests_odd}: $s\in(-10^4,10^4)$, $R = -10$):
\begin{widetext}
\be
r = \sqrt{\frac{-2 \left(4 R^{2} s^{2}-4 R +1\right) \left(\mp \sqrt{4 R^{2} s^{2}-4 R +1}+2 R -1\right) R^{2}}{\left(2 R^{2} s^{2}\mp 2 \sqrt{4 R^{2} s^{2}-4 R +1}\, R -4 R \pm \sqrt{4 R^{2} s^{2}-4 R +1}+1\right)^{2}}},
z = \frac{\mp 2 R^{2} s}{\left(1-2 R \right) \sqrt{4 R^{2} s^{2}-4 R +1}+2 R^{2} s^{2}-4 R +1}
\ee
\end{widetext}

Elliptic case $\nu_4 = \nu_0/2$ (second panel from the left in figure\ref{wave_crests_odd}: $s\in(-1.23,1.23)$, $R = -3$):
\be
r =\frac{1}{2}\sqrt{\frac{R^{2} \left(2 R -1\right) \left(R^{2} s^{2}+2 R \mp 2 \sqrt{-2 R +1}-2\right)}{\left(\pm \left(R -1\right) \sqrt{-2 R +1}+2 R -1\right)^{2}}}
\ee
\be
z = -\frac{\sqrt{-2 R +1}\, R^{2} s}{\left(2 R -2\right) \sqrt{-2 R +1}\pm (4 R -2)}
\ee

Parabolic case $\nu_4 = \nu_0$ (second panel from the right in figure\ref{wave_crests_odd}: $s\in(-10^3,10^3)$, $R = 0.2$):
\be
r = \sqrt{\frac{1+s^{4} R^{2}+\left(2 R -1\right) s^{2}}{\left(s^{4} R^{2}-1\right)^{2}}}, \quad 
z = \frac{s \left(2 R^{2} s^{2}+2 R -1\right)}{s^{4} R^{2}-1}
\ee

Hyperbolic case $\nu_4 =2  \nu_0$ (right-most panel in figure\ref{wave_crests_odd}: $s\in(-10^3,10^3)$, $R = 0.2$):
\begin{widetext}
\be
r = \sqrt{2}\, \sqrt{\frac{R^{2} \left(12 R^{2} s^{2}+4 R +1\right) \left(4 R^{2} s^{2}\pm \sqrt{12 R^{2} s^{2}+4 R +1}+2 R +1\right)}{\left(h(s; R)\right)^{2}}},
z = 
-\frac{2 R^{2} s \left(2 \sqrt{12 R^{2} s^{2}+4 R +1}\pm 3\right)}{\pm h(s; R))}
\ee
\end{widetext}
where $ h(s; R) = \pm 6 R^{2} s^{2}+2 \sqrt{12 R^{2} s^{2}+4 R +1}\, R +\sqrt{12 R^{2} s^{2}+4 R +1}\pm(4 R +1)$.

The $\pm$ signs above correspond to the two roots of $\kappa^2$ when solving equation \rr{P}. 

For a two dimensional compressible odd viscous liquid the dimensionless parameter we employ below is 
\be
s = \frac{2\nu_0 U k_y}{c^2}.
\ee
In figure \ref{wave_crests_odd2D} the curves have the following parametric representation $\mathbf{x}(s) = (x(s), y(s))$ where the coordinate functions are 
\be
(x,y) \rightarrow  \frac{U}{\nu_0}(x,y),
\ee
$s\in(10,220)$ and $V = 10$. Thus,
\be
x = \frac{2 V \sqrt{s^{2}+1}\, \sqrt{\frac{2 \sqrt{s^{2}+1}\, V^{2}-2 V^{2}-s^{2}}{V^{2}}}}{s^{2}-2 \sqrt{s^{2}+1}+2}
\ee
and 
\be
y = -\frac{2 s \left(V^{2}-\sqrt{s^{2}+1}\right)}{s^{2}-2 \sqrt{s^{2}+1}+2}
\ee
and $V$ was defined in \rr{V}. 



\bibliography{biblio}

\end{document}